\documentclass[lettersize,journal]{IEEEtran}
\usepackage{amsmath,amsfonts}
\usepackage{mathtools}
\usepackage{amsthm}
\usepackage{array}
\usepackage{textcomp}
\usepackage{stfloats}
\usepackage{url}
\usepackage{microtype}
\usepackage{verbatim}
\usepackage{graphicx}
\usepackage{cite}
\usepackage[colorlinks=true, linkcolor=blue, citecolor=blue, urlcolor=blue]{hyperref}
\usepackage{enumitem}

\usepackage[linesnumbered,lined,algoruled,commentsnumbered]{algorithm2e}

\usepackage{caption}
\usepackage{subcaption}
\usepackage{multirow}
\usepackage{booktabs}
\usepackage[table]{xcolor}
\usepackage{stfloats}
\usepackage{lipsum} % for placeholder text
\usepackage[font=small,skip=0pt]{caption}
\usepackage[export]{adjustbox}

\SetCommentSty{mycommfont}
\theoremstyle{definition}

\usepackage{units}
\newcommand{\nth}[1]{{#1}^{\text{th}}}

\newcommand{\abs}[1]{\left|{#1}\right|}

\newcommand{\RD}[0]{r_{\mathrm{\scalebox{0.5} {RD} }}}

\newcommand{\rf}[0]{r_{\mathrm{\scalebox{0.5} {0}}}}
\newcommand{\au}[0]{\theta_{\mathrm{\scalebox{0.5} {0}}}}
\newcommand{\BD}[0]{r_{\mathrm{\scalebox{0.5} {BD} }}}

\usepackage{acronym}
\input{acronym}

\usepackage{titlesec}
\titlespacing*{\section}{0pt}{1 pt}{1 pt} 
\titlespacing*{\subsection}{0pt}{0.5pt}{.5pt}
\titlespacing*{\subsubsection}{0pt}{0.25pt}{.25pt}
\begin{document}

\title{Near-Field Tapering with Slepian Window: Balancing the Range–Angle Sidelobe Trade-off\\

\author{Ahmed Hussain,~\IEEEmembership{Graduate Student Member, IEEE}, Ahmed Sultan,~\IEEEmembership{Senior Member, IEEE}, Asmaa Abdallah \\ ~\IEEEmembership{Member, IEEE}, Abdulkadir Celik, ~\IEEEmembership{Senior Member, IEEE}, and Ahmed M. Eltawil,~\IEEEmembership{Senior Member, IEEE}

\thanks{The authors are with Computer, Electrical, and Mathematical Sciences and Engineering (CEMSE) Division, King Abdullah University of Science and Technology (KAUST), Thuwal, 23955-6900, KSA.}
}
}

\maketitle

\begin{abstract}
Near-field beamforming enables target discrimination in both range (axial) and angle (lateral) dimensions. Elevated sidelobes along either dimension, however, increase susceptibility to interference and degrade detection performance. Conventional amplitude tapering techniques, designed for far-field scenarios, cannot simultaneously suppress axial and lateral sidelobes in near-field. In this letter, we propose a Slepian-based amplitude tapering approach that maximizes mainlobe energy concentration, achieving significant sidelobe reduction in both dimensions. Numerical results show that the proposed taper improves peak sidelobe suppression by approximately 24~dB in the lateral domain and 10~dB in the axial domain compared to a conventional uniform window.
\end{abstract}

%%%%%%%%%%%%%%%%%%%%%%%%%%%%%%%%%%%%%%%%%%%%%%%%%%%%%%%%%%%%%%%%%%%%%%%%%%%%%%%%%%%%%%%%%%%%%%%%%%%%%%%%%%%%%%%%%%%%%%%%%%%%%%%%%%
\section{Introduction}
\IEEEPARstart{S}{ensing}, formally recognized as a key service in 3GPP release~20, is envisioned as a core capability of \ac{6G} wireless networks. This paradigm shift is driven by the adoption of \ac{mmWave} bands, coupled with \ac{UM}-\ac{MIMO} architectures, that jointly enable fine spatial resolution through wide bandwidths and large antenna apertures, respectively~\cite{11095387}. At such scales, electromagnetic propagation occurs in the radiative \ac{NF} regime, where spherical wavefronts enable the formation of \ac{NF} beams. These beams exhibit finite beamwidth and beamdepth, thereby facilitating target resolution in both angular (lateral) and range (axial) domains. However, unlike the \ac{FF}, where interference arises solely from \acp{LSL}, \ac{NF} beams exhibit both \acp{ASL} and \acp{LSL}, rendering them susceptible to interference in both axial and lateral dimensions.

Sidelobe suppression is critical for reliable sensing and secure communication, as the mainlobe width determines spatial resolution while \acp{SLL} govern susceptibility to interference. Reducing sidelobes enhances sensing performance by mitigating undesired signal leakage from interference sources and improves multiuser communication performance by increasing the achievable sum rate. The overall lateral beam pattern depends on the product of the element pattern and array factor; hence, reducing element sidelobes directly lowers \acp{LSL}. Conversely, the element pattern does not influence the axial beam pattern. In the \ac{FF}, amplitude weights are applied in the spatial domain to reduce \acp{LSL} in the angular domain~\cite{ogurtsov2021review}. In contrast, within the \ac{NF} context, it is essential to investigate how a single set of amplitude weights (in the spatial domain) can simultaneously suppress both \ac{ASL} and \ac{LSL}. Recent findings reveal a fundamental trade-off between \ac{ASL} and \ac{LSL} suppression: optimizing tapering weights for one domain often degrades performance in the other~\cite{ahmed2025NF,1143106}. This inherent coupling raises a critical research question: \textit{Can we design an amplitude distribution that jointly minimizes \ac{ASL} and \ac{LSL}?}

Interestingly, the trade-off between \ac{ASL} and \ac{LSL} suppression mirrors the classical time--frequency uncertainty, in which a signal cannot be simultaneously time-limited and band-limited. Slepian sequences address this by maximizing the energy concentration of a finite-length sequence within a prescribed bandwidth~\cite{slepian1978prolate}. These sequences are mutually orthogonal, and the principal Slepian sequence yields the most effective window function for sidelobe suppression. Drawing on this analogy, the \ac{NF} problem exhibits a similar duality, as range and angle domains are inherently coupled with unavoidable resolution trade-offs. Motivated by this parallel, in this letter we first characterize the \ac{NF} beam patterns across the range--angle domain and quantify the underlying trade-offs. We then propose Slepian-based amplitude tapers for \ac{NF} beamforming that jointly suppress \ac{ASL} and \ac{LSL}, yielding superior performance compared to existing windowing methods.
\section{System Model}
In this section, we present the \ac{NF} beam pattern and specify the metrics to quantify the \acp{SLL}.
\subsection{Channel Model}
We consider a \ac{ULA} equipped with $N$ isotropic antenna elements spaced $d = \lambda/2$. Based on the planar wavefront assumption, the normalized \ac{FF} array response vector $\mathbf{a}(\theta) \in \mathbb{C}^{N \times 1}$ is formulated as
\begin{equation} 
\mathbf{a} (\theta) = \tfrac{1}{\sqrt{N}}\Big[1, e^{-j\nu d\sin(\theta)} ,\dots, e^{-j\nu d(N-1)\sin{(\theta)}}\Big]^\mathsf{T},
\label{eqn-A1}
\end{equation}
 where $\nu = \frac{2\pi}{\lambda}$ is the wavenumber, and $\lambda$ is the wavelength.
On the other hand, the \ac{NF} array response vector $\mathbf{b}(\theta,r) \in \mathbb{C}^{N \times 1}$ based on the spherical wave model is given by
\begin{equation} 
\mathbf{b} (\theta,r) = \tfrac{1}{\sqrt{N}}\Big[e^{-j\nu(r^{(0)} -r)} ,\dots, e^{-j\nu(r^{(N-1)} -r)}\Big]^\mathsf{T},
\label{eqn-A2}
\end{equation}
 where $r$ is the distance between the focal point and the center of the \ac{ULA}, while $r^{(n)}$ is the distance between the focal point and the $\nth{n}$ antenna element. The corresponding phase shift $\nu(r^{(n)}-r)$, obtained via the law of cosines, is expressed as $\nu(r^{(n)} - r) = \tfrac{2\pi}{\lambda} \left( \sqrt{r^2 + n^2 d^2 - 2 r n d \sin(\theta)} - r \right)$. Accordingly, $\nth{n}$ component of the \ac{NF} array response vector can be written as
\begin{equation} 
{b}_n (\theta,r) = \tfrac{1}{\sqrt{N}}e^{-j\nu \sqrt{r^2+ n^2d^2-2r nd \sin(\theta)} - r}.
\label{eqn-A3}
\end{equation}
The above equation can be simplified by using \textit{near field expansion} based on the second-order Taylor expansion as $\sqrt{1+t} \approx 1+\frac{t}{2}-\frac{1}{8}t^2$ where $\sqrt{r^2+ n^2d^2-2r nd \sin(\theta)} - r \approx nd\sin(\theta)- \frac{1}{2r}n^2d^2\cos^2(\theta) $. Hence, ${b}_n (\theta,r)$ in \eqref{eqn-A3} is approximated as
\begin{equation} 
{b}_n (\theta,r) \approx \tfrac{1}{\sqrt{N}}e^{-j\nu\{nd\sin(\theta)- \frac{1}{2r}n^2d^2\cos^2(\theta)\}}.
\label{eqn-A4}
\end{equation}
\subsection{Problem Formulation}
\begin{figure}[t]
\centering
\includegraphics[width=1\columnwidth]{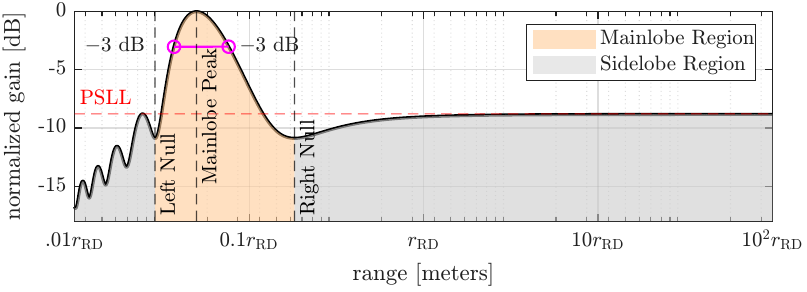}
\caption{\ac{NF} beam pattern with mainlobe and .}
\setlength{\belowcaptionskip}{-25pt}
\setlength{\abovecaptionskip}{0pt}
\label{Range_AF_Shaded}
\end{figure}

The \ac{NF} beam pattern is obtained by applying the complex weights $\mathbf{g}$ to the antenna elements as
\begin{equation}
\mathcal{G}(\theta, r) = \left| \mathbf{g}^\mathsf{H}(\au, \rf) \, \mathbf{b} (\theta, r) \right|^2,
\label{eqn-A5}
\end{equation}
where $(\au, \rf)$ denotes the desired beamfocusing location in angle and range, respectively. The weighting vector $\mathbf{g}$ can be decomposed into element-wise amplitude and phase components as
\begin{equation}
\mathbf{g} = \mathbf{w} \odot e^{j \boldsymbol{\phi}},
\label{eqn-A6}
\end{equation}
where $\mathbf{w}$ is the real-valued amplitude vector, $e^{j \boldsymbol{\phi}}$ is the complex phase vector, and $\odot$ denotes the Hadamard (element-wise) product. In general, the phase component focuses the beam toward the desired location $(\au, \rf)$, while the amplitude component is designed to control the \ac{SLL}. For a given focus point, the phase component $e^{j \boldsymbol{\phi}}$ is obtained using \eqref{eqn-A4}. Substituting the decomposition of $\mathbf{g}$ from~\eqref{eqn-A6} into the beam pattern expression in~\eqref{eqn-A5} yields
\begin{equation}
\mathcal{G}(\theta, r) = \left| \left( \mathbf{w} \odot \mathbf{b}(\au,\rf) \right)^\mathsf{H} \mathbf{b}(\theta, r) \right|^2.
\label{eqn-A7}
\end{equation}
Given the peak of the mainlobe is $\mathcal{G}(\au, \rf)$, the \ac{PSL} and \ac{ISL} are defined as follows:
\begin{enumerate} [label=\alph*)]
    \item  \textbf{\ac{PSL}:} The \ac{PSL} quantifies the ratio of the highest sidelobe magnitude to the mainlobe peak, indicating worst-case sidelobe susceptibility against narrowband interference. 
\begin{equation}
\mathrm{PSLL} = 10 \log_{10} \left( \frac{\max\limits_{(x) \in \mathcal{S}} \mathcal{G}(\theta, r)}{\mathcal{G}(\au, \rf)} \right),
\label{eqn-A8}
\end{equation}
where $x$ denotes either angle ($\theta$) or range ($r$) depending on the domain of interest. $\mathcal{S}$ corresponds to the set of points in the sidelobe region.
\item \textbf{\ac{ISL}:}
The \ac{ISL} quantifies the ratio of the total sidelobe power to the mainlobe power, characterizing the beampattern's susceptibility against wideband interference. 
\begin{equation}
\mathrm{ISLL} = 10 \log_{10} \left( \frac{ \displaystyle \int_{\mathcal{S}} |\mathcal{G}(\theta, r)|^2 \, dx }{ \displaystyle \int_{\mathcal{M}} |\mathcal{G}(\theta, r)|^2 \, dx } \right),
\label{eqn-A9}
\end{equation}
 where $\mathcal{M}$ corresponds to the mainlobe. Fig. \ref{Range_AF_Shaded} illustrates the mainlobe and sidelobe regions, as well as the \ac{PSL}, for a \ac{NF} beam pattern in the range domain. Conventional window designs entail a trade-off between \ac{ASL} and \ac{LSL} suppression, as detailed in Section~\ref{sec-IIID}. To address this limitation, we propose a window that jointly attenuates both \ac{ASL} and \ac{LSL} in terms of \ac{PSL} and \ac{ISL}.
\end{enumerate}
\section{Beam Pattern Analysis} \label{Sec-III}
In this section, we derive the beamwidth, beamdepth, and \ac{PSL} of the \ac{NF} beam in both the angular and range domains, assuming uniform amplitude excitation $\mathbf{w} = \mathbf{1}$. We then highlight the inherent range--angle trade-off in sidelobe suppression.
\subsection{Angle-Domain} \label{Sec 3-A} 
To evaluate the beam pattern in the angular domain, we consider a distance ring defined by $\frac{\cos^2 (\theta)}{r} = \frac{\cos^2 (\au)}{\rf}$ \cite{11078749}. This curve corresponds to a set of range-angle pairs over which the beamwidth remains approximately invariant. Accordingly, the beam pattern in the angle domain can be expressed as 
\begin{equation}\small
\begin{aligned}
\mathcal{G}(\theta) 
&= \left|\mathbf{b}^{\mathsf{H}}(\au,\rf)\mathbf{b}(\theta,r)\right|^2, \\
&\approx \left|\frac{1}{N}\sum_{n=0}^{N-1}
e^{\,j \nu nd (\sin\theta-\sin\au)}\,
e^{-j \nu \tfrac{n^2 d^2}{2}\left(\tfrac{\cos^2\theta}{r}-\tfrac{\cos^2\au}{\rf}\right)}\right|^2, \\
&\stackrel{(a)}{=}
\left|\frac{1}{N}\sum_{n=0}^{N-1}
e^{\,j \nu nd (\sin\theta-\sin\au)}\right|^2, \\
&= \left|\frac{1}{N}\,\frac{\sin\!\big(\tfrac{N \nu d}{2}(\sin\theta-\sin\au)\big)}
{\sin\!\big(\tfrac{\nu d}{2}(\sin\theta-\sin\au)\big)}\right|^2,
\end{aligned}
\label{eqn-B2-11}
\end{equation}
\normalsize
where (a) follows from the relation $\tfrac{\cos^{2}(\theta)}{r} = \tfrac{\cos^{2}(\au)}{\rf}$. 
From \eqref{eqn-B2-11}, \ac{HPBW} $\Delta \theta_{3\text{dB}}$ is derived as 
\begin{equation}
\Delta \theta_{3\text{dB}} \approx \frac{0.886 \lambda}{N d \cos\au}, 
\label{eqn-B5-14}
\end{equation}
which is identical to the \ac{HPBW} expression in the \ac{FF} \cite{van2004optimum}. The \ac{PSL} of a uniformly weighted array is determined by the secondary maxima of the gain function $\mathcal{G} = \abs{\tfrac{\sin \!\left( \tfrac{N}{2} x \right)}{N \sin(x/2)}}^2$ in \eqref{eqn-B2-11}, where $x= \nu (\sin\theta-\sin\au) d$. The mainlobe attains its peak value of unity at $x=0$, while the first sidelobe occurs at $x \approx 3\pi/N$, yielding $\mathrm{PSLL} \approx 10 \log_{10}\!\big(1/(3\pi/2)^2\big) \approx -13.46~\mathrm{dB}$, where a \textit{small angle} approximation is applied in the denominator to obtain $\sin(\frac{3\pi}{2N})\approx \frac{3\pi}{2N}$.
\subsection{Range-Domain} \label{Sec 3-B}
The beam pattern in the range domain is obtained as the inner product of \ac{NF} array response vectors pointing to the same angle $\theta$ but different distances $r$ and $\rf$ \cite{10988573}
\begin{align}
\mathcal{G}(\theta,r) &= \left| \mathbf{b}^\mathsf{H} (\theta, \rf) \, \mathbf{b} (\theta, r) \right|^2, \label{eqn-B6-15}\\
\overset{(\mathrm{a})}{=} \ & \left| \frac{1}{N} \sum_{n=-N/2}^{N/2} e^{ -j\nu n^2 d^2 \cos^2(\theta) r_\mathrm{eff} } \right|^2, \label{eqn-B6-16}\\
\overset{(\mathrm{b})}{\approx} \ & \frac{C^2(\gamma) + S^2(\gamma)}{\gamma^2}, \label{eqn-B6-17}
\end{align}
where $r_\mathrm{eff} = \abs{\frac{r - \rf}{2r\rf}} $ in (a). By introducing the transformation $x = \sqrt{\frac{n^2d^2\cos^2{(\theta)}}{\lambda} \abs{ \tfrac{r -\rf}{r\rf} }} $, $\mathcal{G}(\theta,r)$ is expressed in terms of Fresnel functions in (b), where
$\gamma = \sqrt{\frac{N^2 d^2 \cos^2(\theta)}{\lambda} r_\mathrm{eff}}$. Furthermore, $C(\gamma) = \int_{0}^{\gamma} \cos\left( \tfrac{\pi}{2}x^2 \right) dx$, $S(\gamma) = \int_{0}^{\gamma} \sin\left( \tfrac{\pi}{2}x^2 \right) dx$ are the Fresnel cosine and sine integrals.

\Ac{HPBD} $\BD$ is defined as the distance interval $r \in [\rf^\mathrm{min}, \rf^\mathrm{max}]$ where normalized array gain is at most $\unit[3]{dB}$ lower than its maximum value. For a \ac{ULA}, \cite{10934779}
\begin{align}
\rf^\mathrm{min} &= \frac{\rf \RD \cos^2{(\theta)}}{\RD \cos^2{(\theta)} + 4 \rf \alpha_{3\mathrm{dB}}}, \label{eqn-B7-18a}\\
\rf^\mathrm{max} &= \frac{\rf \RD \cos^2{(\theta)}}{\RD \cos^2{(\theta)} - 4 \rf \alpha_{3\mathrm{dB}}},
\label{eqn-B7-18b}
\end{align}
where $\RD = \frac{2D^2}{\lambda}$ is the Rayleigh distance and $D$ represents the aperture length of the \ac{ULA}. Based on \eqref{eqn-B7-18a} and \eqref{eqn-B7-18b}, $\BD = \rf^\mathrm{max} - \rf^\mathrm{min}$ is given by
\begin{equation} 
\BD = \begin{cases}\frac{8 \alpha_\mathrm{3dB} \rf^2{\RD} \cos^2{(\theta)}} {( \RD \cos^2{(\theta))^2 -(4\alpha_\mathrm{3dB} \rf )^2}},& \rf<\frac{\RD}{4 \alpha_\mathrm{3dB}} \cos^2{(\theta)},\\ \infty, & \rf \geq \frac{\RD} {4 \alpha_\mathrm{3dB}}\cos^2{(\theta)}, \end{cases}
\label{eqn-B7-18}
\end{equation}
where $\alpha_{\mathrm{\scalebox{0.5}{3dB}}} \stackrel{\Delta}{=}\left\{ \gamma \mid \mathcal{G}(\theta,r) = 0.5 \right\}$ denotes the value of $\gamma$, where $\mathcal{G}(\theta,r)$ reduces to half. Furthermore, $\frac{\RD} {4 \alpha_\mathrm{3dB}}\cos^2{(\theta)}$ defines the maximum range limit beyond which $\BD$ goes to infinity. 

To find the \ac{PSL} in the range domain, we solve $\frac{d}{d\gamma} \left( \frac{C^2(\gamma) + S^2(\gamma)}{\gamma^2} \right)$. Solving numerically, the peak of the first sidelobe occurs at $\gamma_{\text{PSLL}} \approx 2.28$, with $\mathcal{G}(\gamma_{\text{PSLL}}) \approx 0.1323$, giving a \ac{PSL} of $10 \log_{10}(0.1323) = \unit[-8.7]{dB}$, which is also highlighted as red dotted line in Fig.~\ref{Range_AF_Shaded}.
 
\begin{table}[t]
\centering
\scriptsize
\caption{Angular vs. range domain beam pattern.}
\renewcommand{\arraystretch}{1}
\setlength{\tabcolsep}{3pt}
\begin{tabular}{|m{2cm}|m{3cm}|m{3cm}|}
\hline
\textbf{Property} & \textbf{Angular Domain} & \textbf{Range Domain} \\
\hline
Beam Pattern & Sinc & Fresnel \\
\hline
Symmetry & Symmetric about $\au$ & Asymmetric about $\rf$ \\
\hline
\ac{PSL} & $\unit[-13.46]{dB}$ & $\unit[-8.7]{dB}$ \\
\hline
Grating Lobes & Present (sinc periodicity) & Absent (non-periodic) \\
\hline
\end{tabular}
\label{Table_1}
\vspace{-15 pt}
\end{table}

\subsection{Range vs. Angle Beam Pattern}
The key differences between the axial and lateral beam patterns are summarized in Table~\ref{Table_1}. Notably, the angular pattern is symmetric about the steering angle $\au$, whereas the range pattern is asymmetric, as observed from \eqref{eqn-B7-18a} and \eqref{eqn-B7-18b}, with $\left|\rf^\mathrm{max} - \rf\right| > \left|\rf - \rf^\mathrm{min}\right|$. As derived in Sections~\ref{Sec 3-A} and~\ref{Sec 3-B}, the \ac{PSL} in the range domain is approximately $\unit[5]{dB}$ higher than that in the angular domain. Furthermore, the periodicity of the sinc function introduces grating lobes in the angular domain when $d > \frac{\lambda}{2}$. In contrast, the Fresnel function is aperiodic, and therefore the range pattern does not exhibit any grating lobes.
\subsection{Range-Angle Trade-off} \label{sec-IIID}
The \ac{FF} beam provides lateral resolution, i.e., the ability to distinguish targets at the same range but different angles. The \ac{NF} beam additionally enables axial resolution, i.e., the ability to distinguish targets at the same angle but different ranges. At short ranges, the beamdepth becomes extremely narrow, whereas the Fourier transform-based angular resolution degrades significantly. Conversely, at larger distances or wider angles, the lateral resolution improves while the axial resolution deteriorates due to the increased beamdepth~\cite[Fig.~1]{10934779}. This trade-off is analogous to the time--frequency uncertainty, which prevents a function from being simultaneously localized in both domains. Consequently, conventional window functions cannot achieve simultaneous sidelobe suppression across both dimensions.

\begin{figure}[!t]
\centering
\begin{subfigure}[t]{0.48\columnwidth}
 \centering
\includegraphics[width=\textwidth]{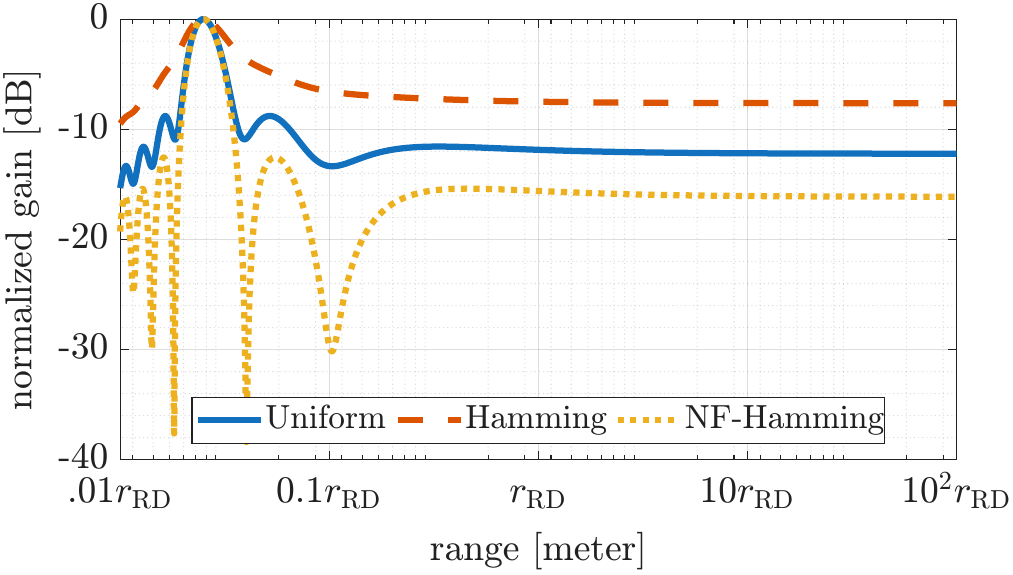}
\caption{}
\label{2a}
\end{subfigure}
 \hfill
 \begin{subfigure}[t]{0.48\columnwidth}
 \centering
 \includegraphics[width=\textwidth]{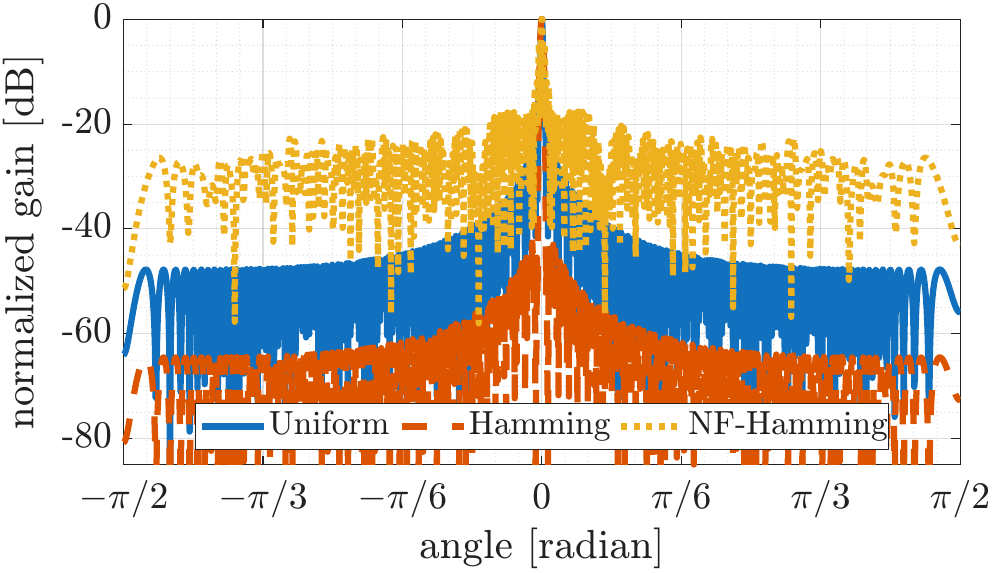}
\caption{}
 \label{2b}
\end{subfigure}
\setlength{\belowcaptionskip}{-20pt}
\setlength{\abovecaptionskip}{0pt}
\caption{(a) The \ac{NF}-Hamming window effectively suppresses the \acp{ASL}, while (b) the conventional Hamming window effectively suppresses the \acp{LSL}.}
 \label{fig2_SLL_tradeoff}
\end{figure}

Standard \ac{FF} techniques apply amplitude tapering as $\mathbf{a}(\theta) \odot \mathbf{w}(n)$, where $\mathbf{w}(n)$ suppresses only the \ac{LSL}. To address this limitation, a transformation is proposed in~\cite{ahmed2025NF}, which maps a conventional window $\mathbf{w}(n)$ to a modified window $\bar{\mathbf{w}}(n)$, thereby enabling existing designs to be directly adapted for \ac{ASL} suppression
\begin{equation}
\bar{\mathbf{w}}(n) = |n|\,\mathbf{w}(n^2).
\label{eqn-B11-22}
\end{equation}
To illustrate further, Fig.~\ref{fig2_SLL_tradeoff} compares \ac{ASL} and \ac{LSL} for three window functions: a uniform (untapered) case, a conventional Hamming window $\mathbf{w}(n)$, and its NF-Hamming $\bar{\mathbf{w}}(n)$ obtained from \eqref{eqn-B11-22}. As observed, the conventional Hamming window effectively suppresses \ac{LSL} but increases \ac{ASL}, whereas the NF-Hamming window reverses this trade-off by enhancing \ac{ASL} suppression at the cost of higher \ac{LSL}, even exceeding that of the uniform window.
\section{Proposed Slepian-Based Window Function} \label{Sec-IV}
In this section, we first review the classical Slepian window and then present the proposed window design.
\subsection{Primer on Slepian Window}
Slepian window achieves optimal trade-off between time and frequency resolutions for a given time--bandwidth product ($NW$) by maximizing the fraction of their total spectral energy within a specified band $[-W, W]$, where $0 < W < f_s/2$ and $f_s$ is the sampling frequency. Let $s(n)$ denote the Slepian sequence in the time domain, and $S(f)$ its Fourier transform, given by $S(f) = \sum_{n=0}^{N-1} s(n) e^{-j 2 \pi f n T_s}$, where $T_s = 1/f_s$. The objective is to determine $s(n)$ that maximizes the energy concentration ratio given by \cite{10988573}
\begin{equation}
\Lambda = \frac{ \int_{-W}^{W} |S(f)|^2 df }{ \int_{-f_s/2}^{f_s/2} |S(f)|^2 df },
\label{eqn-C1-23}
\end{equation}
where $0<\Lambda<1$ is the concentration parameter. The above equation can be solved using Parseval’s theorem to yield \cite{slepian1978prolate}
\begin{equation}
 \Lambda = 
 \frac{ \displaystyle \sum_{n=0}^{N-1} \sum_{m=0}^{N-1} s(n) s(m) A_{n,m} }
{ \displaystyle \sum_{n=0}^{N-1} \lvert s(n) \rvert^2 }
 = \frac{ \mathbf{s}^{\mathsf{T}} \mathbf{A} \mathbf{s} }{ \mathbf{s}^{\mathsf{T}} \mathbf{s} },
 \label{eqn-C1-24}
\end{equation}
where $A_{n,m} = \frac{2W}{f_s} \, \mathrm{sinc}\!\left( \frac{2W}{f_s}(n-m) \right)$. The Rayleigh quotient in \eqref{eqn-C1-24} attains its maximum when $\mathbf{s}$ is the eigenvector associated with the largest eigenvalue of $\mathbf{A}$. This dominant eigenvector achieves maximal spectral concentration within the specified band $[-W, W]$ and serves as the Slepian window.

\subsection{Proposed Slepian-based Window}
 We propose a Slepian-inspired optimization framework based on energy concentration within the desired mainlobe region in the \ac{NF}. This approach generalizes the classical Slepian concentration problem in \eqref{eqn-C1-23} to a two-dimensional range-angle domain, yielding a solution expressed as a generalized Rayleigh quotient.

Let us consider a \ac{NF} beam focused at $(\au, \rf)$, characterized by a finite beamwidth and beamdepth. We define the limits for the mainlobe power and the total power of the beam as follows: the \textit{mainlobe power} is assumed to be primarily concentrated within the angular region $-\Delta \theta_{3\text{dB}}/2 \leq \theta \leq \Delta \theta_{3\text{dB}}/2$ and the range interval $\rf^\mathrm{min} \leq r \leq \rf^\mathrm{max}$, where $\rf^\mathrm{min}$ and $\rf^\mathrm{max}$ denote the beamdepth limits defined in \eqref{eqn-B7-18a} and \eqref{eqn-B7-18b}, respectively, and $\Delta \theta_{3\text{dB}}$ is the 3dB beamwidth defined in \eqref{eqn-B5-14}. Furthermore, the \textit{total beam power} in the range domain is confined within $r \in [r^\mathrm{min},\; r^\mathrm{max}]$, where $r^\mathrm{min} = 0.62 \sqrt{D^3 / \lambda}$ denotes the inner boundary of the radiative \ac{NF} region, and $r^\mathrm{max} = \RD$ represents its outer boundary. In the angular domain, the beam is assumed to span $\theta \in [-\pi/2,\; \pi/2]$.

We recast \eqref{eqn-A4} by defining $\Omega = \tfrac{d}{\lambda} \sin(\theta)$, which leads to $\cos^{2}(\theta) = 1 - \tfrac{\lambda^{2} \Omega^{2}}{d^{2}}$. Accordingly, the $\nth{n}$ component of the \ac{NF} array response vector can be expressed as
\begin{equation}
{b}_n(\Omega, r) = \frac{1}{\sqrt{N}} 
e^{ -j 2\pi \left( \Omega n - \frac{n^2 d^2}{2 r \lambda} + \frac{\Omega^2 n^2 \lambda}{2 r} \right) },
\label{eqn-C2-24}
\end{equation}
Let ${w}_n$ denote the $\nth{n}$ element of the desired window function. The Slepian-based concentration formulation, which maximizes the ratio of mainlobe power to total beam power, is given by \footnote{$\star$ denotes conjugate operation.}
\begin{equation}
\mathcal{J} = 
\frac{
\displaystyle \int_{\Omega_{\min}}^{\Omega_{\max}} 
\int_{\rf^{\min}}^{\rf^{\max}} 
\left| \sum_{n=0}^{N-1} w_n^{\ast} \, b_n(\Omega, r) \right|^2 
\, dr \, d\Omega
}{
\displaystyle \int_{-1/2}^{1/2} 
\int_{r^{\min}}^{r^{\max}} 
\left| \sum_{n=0}^{N-1} w_n^{\ast} \, b_n(\Omega, r) \right|^2 
\, dr \, d\Omega
}.
\label{eqn-C3-25}
\end{equation}
where $\Omega_{\mathrm{min}} = 0.5\sin(-\Delta \theta_{3\mathrm{dB}}/2), \ \ \Omega_{\mathrm{max}} = 0.5\sin(\Delta \theta_{3\mathrm{dB}}/2)$ assuming $d=\frac{\lambda}{2}$. The numerator corresponds to power concentration within the desired mainlobe region, while the denominator integrates total power over the total spatial region. The goal is to maximize $\mathcal{J}$, analogous to maximizing spectral energy in a band in the classical Slepian problem. This leads to the following generalized Rayleigh quotient problem:
\begin{equation}
\mathcal{J} = \frac{ \mathbf{w}^{\mathsf{H}} \mathbf{A} \mathbf{w} }{ \mathbf{w}^{\mathsf{H}} \mathbf{B} \mathbf{w} },
\label{eqn-C4-26}
\end{equation}
where $\mathbf{A}$ and $\mathbf{B}$ are $N \times N$ Hermitian matrices. The elements of $\mathbf{A}$ are defined as
\begin{equation}
A_{n,m} = \int_{\Omega_{\mathrm{min}}}^{\Omega_{\mathrm{max}}} \int_{ \rf^\mathrm{min}}^{\rf^\mathrm{max}} {b}_n(\Omega, r) \, {b}_m^{\ast}(\Omega, r) \, dr\, d\Omega,
\label{eqn-C5-27}
\end{equation}
and the elements of $\mathbf{B}$ are given by
\begin{equation}
B_{n,m} = \int_{-1/2}^{1/2} \int_{r^\mathrm{min}}^{r^\mathrm{max}} {b}_n(\Omega, r) \, {b}_m^{\ast}(\Omega, r) \, dr\, d\Omega.
\label{eqn-C6-28}
\end{equation}
The above integrals are intractable to solve; therefore, approximate expressions for computing $\mathbf{A}$ and $\mathbf{B}$ are provided in Appendix \ref{Appendix_A}. The optimal weight vector $\mathbf{w}$ that maximizes the concentration ratio $\mathcal{J}$ in \eqref{eqn-C4-26} is given by the \textit{dominant generalized eigenvector} of the matrix pair $(\mathbf{A}, \mathbf{B})$ \cite{li2015rayleigh}. Specifically, it is the eigenvector associated with the largest generalized eigenvalue $\lambda$, satisfying the generalized eigenvalue problem
\begin{equation}
 \mathbf{A} \mathbf{w} = \lambda \mathbf{B} \mathbf{w}.
 \label{eqn-C7-29}
\end{equation}
When both $\mathbf{A}$ and $\mathbf{B}$ are Hermitian and $\mathbf{B}$ is positive definite, all generalized eigenvalues $\lambda$ are real. Moreover, the eigenvectors $\{\mathbf{v}_i\}$ are mutually orthogonal under the inner product induced by $\mathbf{B}$
\begin{equation}
\mathbf{v}_i^{\mathsf{H}} \mathbf{B} \mathbf{v}_j = 0, \qquad i \neq j.
\end{equation}
The magnitude of the $\nth{n}$ eigenvalue $\lambda_n$ quantifies the energy concentration of the corresponding eigenvector $\mathbf{v}_n$. Therefore, the optimal tapering window is selected as the eigenvector corresponding to the largest eigenvalue
\begin{equation}
\mathbf{w} = \abs{\mathbf{v}_{\max}},
\end{equation}
where $\lvert \mathbf{v}_{\max} \rvert$ denotes the magnitude of the dominant eigenvector. Note that the eigenvectors of the conventional Slepian sequence are real and mutually orthogonal. In contrast, in our case, the eigenvectors may be complex and are orthogonal with respect to the inner product defined by $\mathbf{B}$.                                                                                                                                                                                                                                                                                                                                                                                                                                                                                                                                                                                                                                                                                                                       
\section{Simulation Results}
We consider a $128$-element \ac{ULA} operating at $\unit[15]{GHz}$, with the \ac{NF} beam focused at a range of $\rf = \tfrac{\RD}{100}$ and an angle of $\au = 0$. The \ac{NF} array response vector is tapered using six window functions: the uniform window (no tapering), the conventional Hamming window, the \ac{NF}-Hamming window derived from~\eqref{eqn-B11-22}, and three variants of the proposed Slepian window. In the Slepian window design, the mainlobe limits act as tunable parameters that are adjusted to achieve the desired balance between \ac{ASL} and \ac{LSL}. In the Slepian-1 configuration, the angular $(\Omega_\mathrm{min},\ \ \Omega_\mathrm{max})$ and range limits $(\rf^\mathrm{min}, \ \ \rf^\mathrm{max})$ of the mainlobe in~\eqref{eqn-C5-27} correspond to the $3$~dB beamwidth and beamdepth obtained from~\eqref{eqn-B5-14}, \eqref{eqn-B7-18a}, and \eqref{eqn-B7-18b} respectively. In Slepian-2, the $3$~dB beamwidth and beamdepth are enlarged by factors of $5$ and $50$, respectively, whereas in Slepian-3, these factors are further increased to $10$ and $100$.

\begin{figure}[!t]
 \centering
 \begin{subfigure}[t]{\columnwidth}
 \centering
 \includegraphics[width=0.8\textwidth]{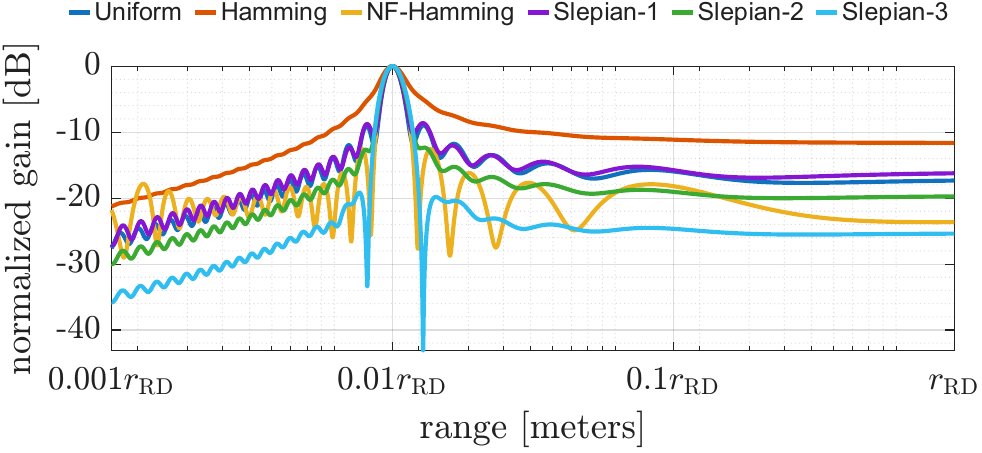}
 \caption{}
 \label{3a}
 \end{subfigure}
 \begin{subfigure}[t]{\columnwidth}
\centering
 \includegraphics[width=0.8\textwidth]{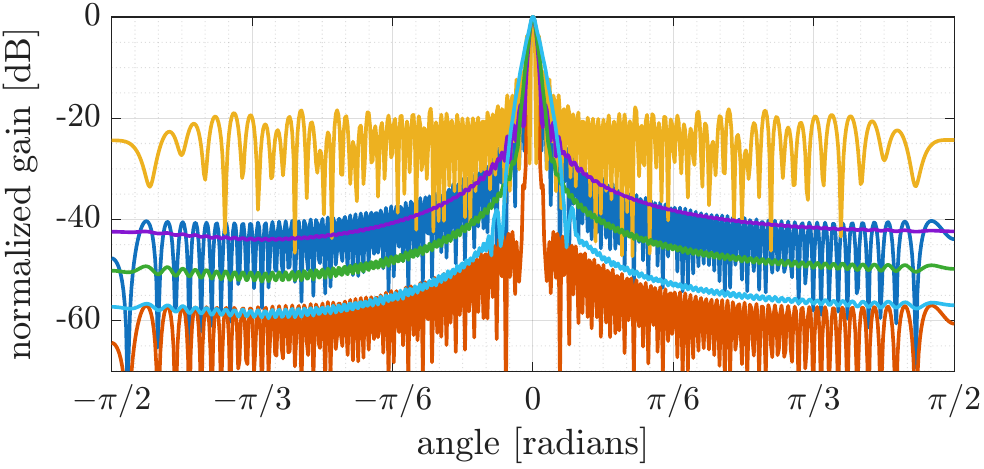}
 \caption{}
 \label{3b}
\end{subfigure}
\caption{Comparison of different windows in suppressing (a) \acp{ASL} in the range domain and (b) \acp{LSL} in the angular domain.}
\label{fig3_Slepian}
\end{figure}

As illustrated in Fig.~\ref{fig3_Slepian}, the Slepian-1 design exhibits behavior similar to the uniform window, as it is explicitly constructed to emulate its characteristics. The Slepian-2 configuration achieves \ac{ASL} and \ac{LSL} suppression levels comparable to those of the NF-Hamming and Hamming windows, respectively. The Slepian-3 window, however, surpasses both the Hamming and NF-Hamming benchmarks in overall performance.

To quantitatively substantiate these observations, Table~\ref{Table_2} summarizes the \ac{PSL} and \ac{ISL} across both range and angular domains. The conventional Hamming window attains $\unit[-33.17]{dB}$ \ac{LSL} suppression but substantially degrades the axial response. Moreover, due to the absence of distinct nulls, identifying the \ac{PSL} becomes challenging. It is also noteworthy that other classical windows exhibit similar degradation in \ac{ASL} suppression; hence, the Hamming window is selected here as a representative case. The \ac{NF}-Hamming window provides $\unit[-12.59]{dB}$ \ac{ASL} suppression, while the \acp{LSL} rise to $\unit[-3.73]{dB}$. Its \ac{ISL} value of $\unit[2.97]{dB}$ further indicates that a considerable portion of the beam energy leaks into the sidelobes rather than remaining confined within the mainlobe.

Both the Slepian-2 and Slepian-3 designs achieve concurrent suppression of \acp{ASL} and \acp{LSL}. Specifically, the Slepian-2 window performs comparably to the Hamming window in the angular domain and to the \ac{NF}-Hamming window in the range domain. In terms of \ac{PSL}, it offers $\unit[5]{dB}$ lower \ac{LSL} suppression than the Hamming window and $\unit[0.27]{dB}$ lower \ac{ASL} suppression compared to the \ac{NF}-Hamming window. The Slepian-3 configuration further enhances the performance of Slepian-2, providing $\unit[4.47]{dB}$ higher \ac{LSL} suppression than the Hamming window and $\unit[6.58]{dB}$ greater \ac{ASL} suppression relative to the \ac{NF}-Hamming. The superior performance of Slepian-3 over Slepian-2, however, comes at the expense of increased beamwidth and beamdepth. This intrinsic trade-off between \ac{SLL} suppression and mainlobe broadening is clearly reflected in Table~\ref{Table_2}. Overall, the proposed Slepian-based tapers achieve a favorable balance by concentrating energy within the mainlobe while effectively suppressing sidelobes in both range and angular dimensions. Moreover, \ac{SLL} attenuation in a specific domain can be flexibly tuned by adjusting the corresponding mainlobe boundaries, as demonstrated by the Slepian-2 and Slepian-3 designs.

\section{Conclusion}
In this paper, we have proposed a Slepian-based amplitude tapering window to suppress \acp{ASL} and \acp{LSL} of the \ac{NF} beam pattern. The proposed design maximizes energy concentration within the mainlobe to reduce the \acp{SLL}. Simulation results demonstrate that the proposed window effectively suppresses both \acp{ASL} and \acp{LSL} by adjusting the mainlobe limits. Future work includes exploring density-based array thinning strategies utilizing the proposed Slepian taper.
\begin{table}[t]
\centering
\renewcommand{\arraystretch}{1} % Optional: add vertical spacing
\caption{Performance comparison of the window functions.}
\begin{tabular}{
 >{\centering\arraybackslash}m{1.6cm} | 
 >{\centering\arraybackslash}m{0.75cm} >{\centering\arraybackslash}m{0.75cm} | 
 >{\centering\arraybackslash}m{0.75cm} >{\centering\arraybackslash}m{0.75cm} | 
 >{\centering\arraybackslash}m{0.75cm} >{\centering\arraybackslash}m{0.75cm}
}
\toprule
\multirow{4}{*}{\textbf{Window}}
& \multicolumn{2}{c|}{\textbf{PSLL [dB]}} 
& \multicolumn{2}{c|}{\textbf{ISLL [dB]}} 
& \multicolumn{2}{c}{\textbf{Mainlobe}} \\
\cmidrule(lr){2-3} \cmidrule(lr){4-5} \cmidrule(lr){6-7}
& \textbf{Range} & \textbf{Angle} 
& \textbf{Range} & \textbf{Angle} 
& \textbf{BD [m]} & \textbf{BW [deg]} \\
\midrule
\centering Uniform & -8.98 & -13.46 & 0.05 & -8.24 & 0.24 & 0.85 \\
\hline
\centering Hamming & NA & \textbf{-33.17} & NA & \textbf{-30.50} & 0.42 & 1.20 \\
\hline
\centering \ac{NF}-Hamming & \textbf{-12.59} & -3.73 & \textbf{-4.26} & 2.97 & 0.27 & 0.85 \\
\hline
\centering Slepian-1 & -8.48 & -17.34 & 0.35 & -12.24 & 0.25 & 1.32 \\
\hline
\centering Slepian-2 & -12.32 & -28.08 & -3.60 & {-22.38} & 0.28 & 1.38 \\
\hline
\centering Slepian-3 & \textbf{-19.17} & \textbf{-37.64} & \textbf{-9.68} & \textbf{-33.09} & 0.29 & 1.84\\
\bottomrule
\end{tabular}
\label{Table_2}
\end{table}

\bibliographystyle{IEEEtran}
\bibliography{IEEEabrv,my2bib}
\appendix
\label{Appendix_A}
Define $\Delta = m-n,\quad \Sigma = m+n,\quad
\alpha = \frac{d^2}{2\lambda},\quad \beta = \frac{\lambda}{2}$. Then, we simplify \eqref{eqn-C5-27} based on \eqref{eqn-C2-24} and above definitions 
\begingroup
\small
\begin{align}
A_{n,m} 
&= \int_{\Omega_{\mathrm{min}}}^{\Omega_{\mathrm{max}}}
e^{\,j 2 \pi \Delta \Omega} 
\int_{\rf^\mathrm{min}}^{\rf^\mathrm{max}}
e^{\,j 2 \pi \frac{\Delta \Sigma}{r} \left(\beta \Omega^2 - \alpha \right)} \,\mathrm{d}r \, \mathrm{d}\Omega \notag\\
&\overset{(a)}{=} \int_{\Omega_{\mathrm{min}}}^{\Omega_{\mathrm{max}}}
e^{\,j 2 \pi \Delta \Omega} 
\left[r\, e^{\,j \frac{C(\Omega)}{r}} -j C(\Omega)\, \operatorname{Ei}\!\Big(j \frac{C(\Omega)}{r}\Big)\right]_{\rf^\mathrm{min}}^{\rf^\mathrm{max}} \mathrm{d}\Omega, \notag
\end{align}
\normalsize
\endgroup
where $C(\Omega) = 2\pi \Delta \Sigma \big(\beta \Omega^2 - \alpha \big)$, and $\operatorname{Ei}(\cdot)$ denotes the exponential integral function. Step (a) follows by applying the identity $\int e^{\,j \frac{C}{r}} \, \mathrm{d}r 
= r\, e^{\,j \frac{C}{r}} - C\, \operatorname{Ei}\!\Big(j \frac{C}{r}\Big)$. The exponential integral term arises due to the nonlinear coupling between $r$ and $\Omega$, making a closed-form evaluation of the double integral intractable. 

Instead we approximate ${A}_{n,m}$ numerically by computing Riemann sum over the angular parameter $\Omega$ and the range parameter $r$ on the grid points $(\Omega_i, r_j)$ within the concentration region defined by $\Omega_i \in [\Omega_{\mathrm{min}}, \  \Omega_{\mathrm{max}})]$ and $r_j \in [\rf^{\mathrm{min}}, \rf^{\mathrm{max}}]$. The summation uses $N_{\Omega}$ samples along the angular ($\Omega$) axis and $N_r$ samples along the range ($r$) axis.

\begin{equation}
{A}_{n,m} \approx \sum_{i=1}^{N_{\Omega}} \sum_{j=1}^{N_r} b_n(\Omega_i, r_j) \, b_m^{\ast}(\Omega_i, r_j)  \Delta \Omega  \Delta r
\label{eqn:RiemannSumA}
\end{equation}
 where $\Delta \Omega$ and $\Delta r$ denote the angular and range step sizes.  Likewise, ${B}_{n,m}$ can be computed
from \eqref{eqn:RiemannSumA} , where the angle–range grid now spans the entire \ac{NF} as defined in \eqref{eqn-C6-28}.
\end{document}